# "Classical-ish": Negotiating the boundary between classical and quantum particles

Benjamin W. Dreyfus[1], Erin Ronayne Sohr[1], Ayush Gupta[1], and Andrew Elby[2]
[1]*Department of Physics, University of Maryland, College Park MD 20742*
[2]*Department of Teaching and Learning, Policy and Leadership,
University of Maryland, College Park MD 20742*

**Abstract.** Quantum mechanics can seem like a departure from everyday experience of the physical world, but constructivist theories assert that learners build new ideas from their existing ones. To explore how students can navigate this tension, we examine video of a focus group completing a tutorial about the "particle in a box." In reasoning about the properties of a quantum particle, the students bring in elements of a classical particle ontology, evidenced by students' language and gestures. This reasoning, however, is modulated by metacognitive moments when the group explicitly considers whether classical intuitions apply to the quantum system. The students find some cases where they can usefully apply classical ideas to quantum physics, and others where they explicitly contrast classical and quantum mechanics. Negotiating this boundary with metacognitive awareness is part of the process of building quantum intuitions. Our data suggest that (some) students bring productive intellectual resources to this negotiation.

## I. INTRODUCTION

Of the many challenges facing learners of quantum mechanics (QM) [1–3], one is that developing a conceptual understanding involves activating new ontologies [4–8]. A student's "ontology" of a given entity [9–11] is what *kind* of entity they think it is, whether literally or metaphorically [12–14]. For instance, in classical physics, protons and sound waves could be seen as two different kinds of entities, corresponding to the ontologies of *matter (or particle)* and *waves* respectively. A quantum "particle" differs from both these kinds of entities; while it possesses attributes of both a classical *particle* and a classical *wave*, it displays other characteristics with no clear classical analogs. Part of conceptual expertise in quantum mechanics is understanding how this new type of entity behaves. In developing this expertise, how might students build on their formal and intuitive physics knowledge, corresponding largely to *particle* and *wave* ontologies?

The relationship between intuitive knowledge about the physical world and formal physics knowledge has been widely explored in physics education research, from various theoretical perspectives. This research, however, focuses mostly on student reasoning about Newtonian mechanics [9,15–17]. This makes sense because Newtonian mechanics deals with phenomena encountered in everyday life. The role of intuitive physics knowledge in learning quantum mechanics is less clear, because quantum phenomena are more removed from everyday experience. Yet, when students learn QM, classical intuitions and associated ontologies, as well as formal classical concepts learned in previous courses, are unavoidably present in students' minds. For this reason, another essential element of developing quantum expertise is metacognition, or thinking about one's own thinking [18]; students must work out when and why they can and cannot rely on classical ontologies and ways of thinking. This aspect of metacognition is *conditional knowledge*, "knowing why and when to do things"[19].

In this preliminary study, we analyze video of students working together on a "particle in a box" tutorial to make sense of a quantum system. Our research questions are, *How do students negotiate the boundaries between classical and quantum ontologies—specifically, in what ways do students can invoke and juxtapose these ontologies? How is metacognition involved in these negotiations?*

## II. METHODOLOGY

Our case-study data come from a video recording of a group of five students working through a new tutorial that we developed on the "particle in a box" (PIAB), in a clinical (non-classroom) setting.. The students, Al, Bob, Chad, Dan, and Ed (pseudonyms), were physics majors enrolled in the first semester of a two-semester upper-level quantum mechanics course. The session took place toward the end of the semester. So the students had previously encountered the PIAB in class and had moved on to more complicated systems, but were now revisiting the PIAB.

For a PIAB (or infinite square well) in one dimension, the potential is zero within a defined range of positions and infinite everywhere else. It is often taught in introductory QM courses, partly because it is one of the few quantum systems for which there is a simple exact analytical solution to the Schrödinger equation. However, behind the mathematical simplicity is conceptual complexity [20,21], stemming in part from the infinite potential (which is usually swept un-der the rug by saying the wavefunction must be zero outside the "box").

The first part of the tutorial asks students conceptual questions about the position, energy, and speed of a quantum PIAB in the ground state. Next, the students are asked to think about these same properties for a classical standing wave on a string, and for a classical particle bouncing around inside an actual box. Finally, students are asked to discuss whether the quantum PIAB is more like a classical particle, a classical wave, both, or neither. In this paper, we analyze the first part of the tutorial, to explore what ontologies and analogies students invoke and negotiate *before* the tutorial prompts them to compare quantum and classical entities.

After viewing the video collaboratively, we selected shorter episodes for deeper analysis based on the presence of links between quantum and classical physics in the students' discussion. We found two episodes from the first part of the tutorial. Both involved students trying to answer "non-standard" questions—ones they likely had not encountered in lecture, textbooks, or homework. In the first clip, the students are addressing the question "Why isn't the ground state $n = 0$? That is, why isn't it possible for the particle to have zero energy?" In the second clip, the students are addressing the question "If you were to measure the speed of the particle at some point in time, what would you expect to measure? Why? Will you get the same measurement every time?" (Unlike position, momentum, and energy, speed is rarely discussed in QM courses.)

Our analysis focuses on several features: metacognitive moments in which students discuss how their thinking about quantum physics relates to their thinking about classical physics; and gestures [22] and descriptive words/phrases [23,24] that can indicate ontologies, according to previous research [24,25]. Some of the observed reasoning is canonically correct, and some is not, but this distinction is not the focus of our analysis.

This analysis methodology is motivated by a dynamic model of ontologies [11,26]. That is, we assume that the ontological categories into which students and experts implicitly place physical concepts are not fixed, but can vary based on context. Students and experts not only switch among multiple ontological categories for the same physical entity, but can also blend multiple ontologies into a single mental space [14]. Because of this, our analysis focuses on what a student (or group of students) is thinking in a particular moment or over a few minutes; we are unable to generalize about a student's stable beliefs.

## III. RESULTS AND DISCUSSION

### A. Episode 1: Why can't a particle have zero energy?

In this episode, the students are responding to question 3, "why isn't it possible for the particle to have zero energy?" After some preliminary discussion, Chad says "If the energy was zero, wouldn't it be like there was no particle in the box anyway? So it's not the same problem? … You can't have a particle with no energy. That's like saying I have a whole bushel of no apples." On the word "bushel," Chad uses both hands to gesture the shape of a (presumably empty) container. For Chad in this moment, the question is trivial, because "zero energy" is simply not a property that a particle can have.

Al responds, explaining why the question is not trivial: "No. What I think they're saying is the difference, like, you could imagine, like in a classical sense, like a ball *[holds out fist to represent a ball]*, in a well. It could just be sitting there *[gestures again with fist, presumably to emphasize that it isn't moving]*. It could have no kinetic energy *[moves fist back and forth]* whatsoever. … What they're saying, I think, is why in a quantum realm it can't—". Al is countering Chad's suggestion by showing that a classical ball could have zero kinetic energy while sitting motionless in a well, and clarifying that the real question is why a quantum PIAB is different from a classical particle (the ball) in this respect. Al explicitly invokes classical mechanics in contrast to quantum mechanics. His gestures are also consistent with this classical particle ontology, representing an object with a fixed position in space.

Addressing the tutorial question the group then takes a journey through linear algebra to ask whether zero can ever be an eigenvalue. Then Al brings up the uncertainty principle and Bob agrees enthusiastically.

**Bob:** *If it has no energy, then, but now it has a definite position, and—*
**Dan:** *And a definite momentum.*
**Bob:** *Yeah! It has a definite position and a definite momentum, which is impossible. 'Cause you know its momentum is zero, and you know its position is right there, which is not possible.*
**Chad:** *Well, do you know where its position in the box is, if it has no momentum? Unless you know the initial state, you don't know[…]*
**Bob:** *But you know it's somewhere in a discrete position, and you know it's, once you find it, it's gonna be right there.*

In this exchange, Bob treats the particle as a hybrid classical/quantum entity. On the one hand, he relies on the implicitly classical attribute that a particle at rest must have a definite position, even if observers don't (yet) know what it is (or, equivalently, that a particle that doesn't have a definite position must not be at rest). On the other hand, he uses this as an input into the uncertainty principle, arguing that the particle cannot have both a definite position and a definite momentum, which is an attribute of a quantum but not a classical particle. Putting these classical and quantum ideas together, he comes up with a proof by contradiction for why the particle cannot have zero energy.

### B. Episode 2: Does the speed change?

The next questions in the tutorial ask what you would expect to measure if you measure the particle's position and energy, and the group answers these without much difficulty. But then they are taken aback by question 6 on speed, perhaps because their quantum course has never addressed "speed." They first address this by translating "speed" into "momentum," a treatment of which they have seen in their course. Chad, however, is unsatisfied by this, asking "Does the particle actually have speed?"

**Chad:** *It's particle in a box. Just bouncing off the walls back and forth?* [slowly moves his pencil back and forth to represent the particle] *Especially if we're just saying it's one-dimensional. Does it slow down at the edges?*
**Dan:** *Well, it has to. Velocity has to change.*
**Al:** *Well, no, it doesn't, because the potential's constant.*
**Ed:** *It's just two walls* [gestures with both hands to form vertical walls]*, it's not—* [uses one hand to show a sloped wall].
**Al:** *Right. So it's not like harmonic* [moves pencil back and forth in the shape of a parabola, representing a harmonic oscillator potential, or perhaps a particle moving in a harmonic oscillator potential]*, it's not slowing down, it's going at the same speed* [moves pencil back and forth in a straight line, representing the particle bouncing back and forth].
**Dan:** *Velocity has to switch direction.*
**Chad:** *The velocity would change its direction.*
**Bob:** *Now we sound like we're switching from quantum to classical explanations.*
**Chad:** *Yeah.* [Everyone laughs] *That's why I was worried about saying that.* [Everyone laughs again]
**Al:** *But what I'm saying is, there's no—if the potential is constant along the bottom of this well, there's no reason why its speed would change.*
**Ed:** *Yeah. It's true.*
**Chad:** *But, because it's infinite, if it's infinite walls, then it can't, like, go into it at all.*
**Al:** *Right.*
**Chad:** *So that disregards all quantumness. All of like the—* [gestures a decaying exponential function with one hand].
**Al:** *There's no like leaking into the—* [gestures decaying exponential functions with both hands].
**Chad:** *Yeah. Leaking.*
**Al:** *Forbidden region.*
**Chad:** *So it's pretty much classical, isn't it.*
**Bob:** *I'd say it's pretty much classical. It's just—*
**Al:** *Well—*
**Chad:** *Classical-ish. 'Cause it has no reason to change speed.*

This dialogue is punctuated by explicit tagging of classical and quantum reasoning. In the first portion, the students use gestures and descriptive phrases consistent with a classical particle ontology: Chad says "bouncing off the walls," and both Chad and Al represent a particle bouncing back and forth with their gestures. Several students use the potential well metaphor [12] to discuss the "well," "walls," and "edges" as if they are physical objects. This line of reasoning, grounded in classical notions of box and particle, appears to be generative: Chad first moves his pencil back and forth to act out a bouncing particle, and this seems to lead him to the question "Does it slow down at the edges?"

Bob shifts the conversation by forcing everyone to go "meta" when he says "Now we sound like we're switching from quantum to classical explanations." Chad initially reacts as if he has been caught red-handed, demonstrating this by grinning and leaning back. But upon further reflection, Chad defends his original explanations. In this moment, Chad sees "leaking" (tunneling) as the marker of "quantumness"; Chad and Al gesture the shape of a quantum wavefunction that leaks through a potential barrier. This use of gestures to illustrate what the system is **not** resembles Al's fist gesture in episode 1, which represented a classical particle that he claimed the PIAB is **not**.) Because there is no leaking through the infinite potential barrier, Chad convinces himself (and Bob) that the system is "pretty much classical," which he qualifies (upon Al's protest) to "classical-ish." (We cannot tell if the "-ish" disclaimer expresses conceptual content or serves as a discursive move to forge consensus.)

In summary, between these two episodes, we see students exhibiting multiple stances towards the nature of the PIAB: the PIAB is a quantum particle, not a classical particle; the PIAB is a classical particle, not a quantum particle; we can explain the PIAB with a combination of classical and quantum properties.

## IV. CONCLUSIONS AND IMPLICATIONS

First, this study supports earlier arguments [11,26] that students are capable of switching among multiple ontologies when reasoning about a given physical entity. In these two brief episodes, we see the students sometimes treating the PIAB as a classical-ish particle bouncing between two walls, sometimes as a quantum entity that cannot have a definite position and momentum at the same time, and sometimes as a "mixed" entity with both classical and quantum properties. (Other groups in our larger data corpus displayed similar flexibility.) For instructors who think that sophisticated understanding of QM includes ways of thinking drawn from classical particle and waves combined with new "quantum" ideas, students' ontological flexibility is an intellectual resource on which instruction can draw and build.

Second, and perhaps more strikingly, we see students spontaneously engaging in metacognition as they negotiate the boundary between classical and quantum entities, trying to figure out when their classical intuitions do and do not apply. This type of "meta" reasoning, which involves thinking about when quantum systems are analogous to the corresponding classical systems and when the unique aspects of quantum mechanics come into play, is already well-known to experts. The fact that some students can engage in this reasoning spontaneously, before a tutorial prompts them to do so, does *not* indicate that all students would get there on their own, or even that the same students would get there in all contexts. However, in PER, it often turns out that what some students can do spontaneously, most students can do with proper scaffolding. In our future work we seek to explore the instructional impacts of scaffolding to guide students towards the kind of "meta" reasoning about classical and quantum mechanics that was initiated spontaneously by the students in this study.

This preliminary study leaves a number of issues unresolved. We have observed ontological flexibility in reasoning about quantum mechanics, but how can we determine how this affects other aspects of developing conceptual understanding? How can we assess whether students are reasoning metacognitively about quantum mechanics when they are not explicit about it? Furthermore, our analysis has not addressed students' larger-scale "philosophical" views about the interpretation of quantum mechanics [27]; how do these stances interface with the finer-grained ontological dynamics exhibited when students reason about specific quantum scenarios? These questions will be appropriate subjects for future work.

## ACKNOWLEDGMENTS

The authors thank the University of Maryland Physics Education Research Group, and our collaborators at the University of Colorado (Noah Finkelstein, Katie Hinko, Jessica Hoy, and Doyle Woody). We also thank Carolina Alvarado, Jesper Haglund, Gina Passante, and Rachel Scherr for constructive feedback through the FFPER GS/P Symposium. This work is supported by NSF-DUE 1323129.